# Carrier Mobility and High-Field Velocity in 2D Transition Metal Dichalcogenides: Degeneracy and Screening

**José M. Iglesias**[1,*], **Alejandra Nardone**[1], **Raúl Rengel**[1], **Karol Kalna**[2], **María J. Martín**[1], **and Elena Pascual**[1,**]

[1]Department of Applied Physics, University of Salamanca, Salamanca E-37008, Spain
[2]Nanoelectronic Devices Computational Group, Faculty of Science & Engineering, Swansea University, Swansea, SA1 8EN, Wales, United Kingdom

E-mail: *josem88@usal.es, **elenapc@usal.es

**Abstract.** The effect of degeneracy and the impact of free-carrier screening on a low-field mobility and a high-field drift velocity in $MoS_2$ and $WS_2$ are explored using an in-house ensemble Monte Carlo simulator. Electron low field mobility increases to 8400 cm$^2$/Vs for $MoS_2$ and to 12040 cm$^2$/Vs for $WS_2$ when temperature decreases to 77 K and carrier concentration is around $5\times10^{12}$ cm$^{-2}$. In the case of holes, best mobility values were 9320 cm$^2$/Vs and 13290 cm$^2$/Vs, reached at 77 K, while at room temperature these fall to 80 cm$^2$/Vs and 150 cm$^2$/Vs for $MoS_2$ and $WS_2$, respectively. The carrier screening effect plays a major role at low fields, and low and intermediate temperatures, where a combination of large occupancy of primary valleys and carrier-phonon interactions dominated by relatively low energy exchange processes results in an enhanced screening of intrinsic scattering. For electrons, degeneracy yields to transport in secondary valleys, which plays an important role in the decrease of the low field mobility at high concentrations and/or at room temperature. The high-field drift velocity is not much affected by carrier screening because of an increased carrier scattering with surface optical polar phonons, favouring larger phonon wavevector interactions with small dielectric function values.

## 1. Introduction

Transition metal dichalcogenides (TMDs) and, specifically, their atomically thin version, are in the spotlight due to their promising electronic, optical, and mechanical properties. When combined with a direct bandgap, TMDs hold a great promise for a potential development of applications in areas such as electronics, optoelectronics, spintronics, energy, and sensing [1]. These monolayer materials have multiple applications as electronic devices such as multifunctional diodes [2], transistors [3, 4], photodetectors [5, 6], solar cells [7, 8], flexible electronics [9, 10], or biosensors [11], just to cite some. Besides, there is a recent and increasing interest in research of new TMD based 2D materials with different electronic properties, such as diverse approaches to $MA_2Z_4$ monolayers due to its geometry and electronic properties, as shown in [12] for $MoSi_2P_4$ monolayers, for example.

A key issue for the future of TMD device technology is an accurate knowledge of the carrier mobility and the high-field drift velocity, which are fundamental transport properties of TMDs. Significant deviations between experimental and modelling works are still found nowadays, mainly due to the large surface-to-volume ratio of atomically thin TMDs, which yields an important sensitivity to environmental factors [13].

While theoretical models have predicted an intrinsic room temperature electron mobility near 410 cm$^2$/Vs for atomically thin $MoS_2$ [14] or 1100 cm$^2$/Vs for $WS_2$ [15], experimental values are usually much lower. A room temperature mobility of 83 cm$^2$/Vs in a monolayer $MoS_2$ transistor has been recently achieved by applying electron-beam irradiation [16]. A record mobility of 33 cm$^2$/Vs is claimed for $WS_2$, while the best $MoS_2$ mobility of 47 cm$^2$/Vs was extracted from data from 390 fabricated FET devices [17]. High density of traps and charged impurities have been identified as major sources of transport degradation in TMDs [3, 18, 19, 20]. The use of substrates with a high dielectric constant [18] or an encapsulation within hexagonal boron nitride [21] have been presented as solutions to attain impurity screening. The influence of the gate bias in double-gated FETs has been studied for accomplishing the reduction of the effective traps [20]. Until experimental fabrication methods for TMDs reach a more mature level, an accurate modelling of transport in these materials becomes critical to guide the experimental efforts. In particular, the influence of the dielectric environment and the screening effects, together with the variation of the lattice temperature, are extremely important [18]. However, in the existing literature, the influence of secondary valleys of the conduction band in TMDs is frequently neglected [18, 14]. Other models consider upper valleys, but a thorough treatment of degeneracy and screening is frequently disregarded [22, 23]. Various works [18, 24] have reported on the influence of the environment (i.e., top and bottom substrates) and carrier density on the $MoS_2$ electron transport characteristics, finding that the surrounding dielectrics broadly limit carrier mobility, and finding a mobility enhancing effect of the carrier density due to free electron screening. Yet, a detailed study of the interplay of screening and temperature on the different scattering mechanisms is lacking. A recently presented Monte Carlo study of mobility in $MoS_2$ has considered scattering with Coulomb centres (the ionised impurity scattering), neutral defects, and surface optical phonons, in addition to the electron scattering with intrinsic phonons [25]. However, static screening was only partially considered for Coulomb centers and neutral defects. .

In this work, we use our in-house ensemble Monte Carlo (EMC) simulator to study the effect of degeneracy and screening on a low-field electron mobility and a high-field drift velocity, focusing on the dependence of both quantities on carrier concentration and temperature. A full carrier screening of scattering events, including intrinsic processes, has been taken into account [18]. The results show that free carrier screening, along with valley occupation and different probability dependence of scattering mechanisms on concentration and temperature are the key to understand non-monotonic behavior of the mobility as a function of a carrier concentration in the most common TMDs, MoS$_2$ and WS$_2$.

## 2. Ensemble Monte Carlo Model

The results presented in this work have been obtained by means of an in-house ensemble Monte Carlo (EMC) simulator. The simulator was successfully tested in the past for different 2D materials such as graphene [26], silicene [27], and various TMDs [28, 29]. The transport model features a multi-band, multi-valley band structure. The conduction band of the TMD materials is described by primary valleys (K points of the first Brillouin zone) and secondary valleys (Q points) using parabolic dispersion relations close to the valley minima. In the valence band, the maxima are also located in the K points, as for direct gap materials, while the secondary valleys lie at the Γ point at lower energy (see Table 1). The effective masses for electrons and holes are extracted from density functional theory (DFT) calculations [22, 30]. In the case of the K valleys, isotropic masses are considered, while for the Q valleys, longitudinal and transversal effective electron masses are taken into account. The values of the effective masses are gathered in Table 1. This analytical description for the bands has shown a good agreement with full-band models in Monte Carlo simulations for TMDs, and being more efficient from the computational point of view [25].

| TMD | $\epsilon^c_{0,Q} - \epsilon^c_{0,K}$ (meV) | $\epsilon^v_{0,\Gamma} - \epsilon^v_{0,K}$ (meV) | $m^c_K$ ($m_0$) | $m^c_{Q,\parallel}, m^c_{Q,\perp}$ ($m_0$) | $m^v_K$ ($m_0$) | $m^v_\Gamma$ ($m_0$) |
|---|---|---|---|---|---|---|
| MoS$_2$ | 70 | 148 | 0.50 | 0.62, 1.00 | 0.58 | 4.05 |
| WS$_2$ | 67 | 173 | 0.31 | 0.60, 0.60 | 0.42 | 4.07 |

**Table 1.** The difference of potential energy of the K and Q valleys in the conduction band, and of the K and Γ valleys in the valence band, the effective electron masses in the different valleys of the conduction and valence bands. $m_0$ denotes the electron mass in vacuum.

The energy-dependent scattering probability is described using the deformation potential formalism, considering intra- and intervalley acoustical phonon branches, , optical phonon branches, and the scattering with the surface polar optical phonons (SPPs) from the SiO$_2$ substrate, also known as remote phonons. The approximation of adeggregated modes is assumed for transverse and longitudinal acoustic modes, as well as for transverse, longitudinal, and optical branches [22, 30]. The screening of free carriers is also incorporated to evaluate the influence of carrier degeneracy on electronic transport. For this purpose, a feasible approach is the inclusion of the dielectric function $\epsilon(q)$ into the scattering matrix [22]. In our case, $\epsilon(q)$ is described by the modified Lindhard's function [18], that also accounts for the dielectric mismatch between the underlying and top interfaces and the TMD layer. In this way, screening is fully accounted for, including intrinsic phonon interactions. The secondary valleys are also included allowing for an adequate evaluation of degeneracy at high temperature/high fields as opposed to the inclusion of only the primary valleys [18]. Finally, the Pauli exclusion principle is also incorporated in the model by discretization of the reciprocal space and the use of a rejection technique [31] for a final state selection following every scattering event. An exhaustive description of our Monte Carlo model is included in the supplementary material.

## 3. Electron and Hole Mobility and High-Field Transport

The analysis of the electron mobility in MoS$_2$ and WS$_2$ is carried out with respect to temperature and carrier concentration, delving into the microscopic phenomena that affect transport properties at low electric fields (0.5 kV/cm). In the context of first-principles material modelling of carrier transport, several approaches devised to obtain the mobility have been proposed [32] depending on the theoretical framework. The EMC method provides a stochastic and intuitive approximation to the Boltzmann Transport Equation that allows extracting the drift velocity, $\langle v \rangle$ by simply averaging the carriers independent velocities, and also obtaining the diffusion coefficient through the study of velocity fluctuations [33]. Within this framework, the mobility can be obtained either by calculating the slope of the low-field drift velocity-electric field relation, or by using the Einstein relation on the diffusion coefficient. In this work, we obtain the mobility by using the first method, as $\mu = (\langle v \rangle / E)|_{\text{lowfield}}$, using the low-field value of $E = 0.5$ kV/cm. The structure chosen for the study consists of a TMD layer sitting on the top of a SiO$_2$ dielectric substrate, the most common substrate used with 2D materials. The samples are considered to be pristine, and free of impurities, defects or wrinkles, which in previous works [18, 25, 24, 3, 18, 19, 20] have been demonstrated to be some of the largest sources of mobility degradation in TMDs. Since impurities and defects are a result of a still immature stage of the fabrication technology and therefore represent unwanted, –yet in principle, avoidable– sources of scattering, the results shown here must be considered as the best scenario.

Figure 1 (a) and (b) show the dependence of the electron and hole mobility with the carrier density at four different temperatures presenting similar trends for both types of carriers, with larger mobility values for holes than for electrons. A significant drop in the mobility values between 77 K and 300 K is observed. At 77 K, an increase in the mobility is seen for the non-degenerate case up to a concentration value around $n \approx 5 \times 10^{12}$ cm$^{-2}$ for electrons, where the maximum electron mobility occurs for both materials (8400 cm$^2$/Vs for MoS$_2$ and 12040 cm$^2$/Vs for WS$_2$). In the case of holes, the increase of the mobility for the non-degenerate case is observed up to $p \approx 6 \times 10^{12}$ cm$^{-2}$, with maximum values of 9320 cm$^2$/Vs for MoS$_2$ and 13290 cm$^2$/Vs for WS$_2$. As the temperature increases, the maximum becomes less prominent, almost disappearing at room temperature. The mobility drop occurs at large concentrations, being also less significant at room temperature. We also plot the mobility obtained when the carrier screening is excluded from the model (i.e., by setting the dielectric function, $\epsilon(q)$, to 1), in order to assess the relevance of screening. With the effect of screening excluded from the simulation, the mobility does not show a maximum at intermediate concentrations. Instead, a very slight variation is observed up to the range of $10^{12} - 10^{13}$ cm$^{-2}$ for both electrons and holes, followed by a progressive and more noticeable drop at larger concentrations. This difference indicates the reason behind the mobility gain at intermediate concentrations, i.e., the effective screening of electron-phonon interactions. The increase in the mobility is substantially affected by the lattice temperature, becoming less pronounced as the temperature increases. Note that our findings are quantitatively different than what was reported in [18], which will be explained later. This mobility improvement can be explained as follows.

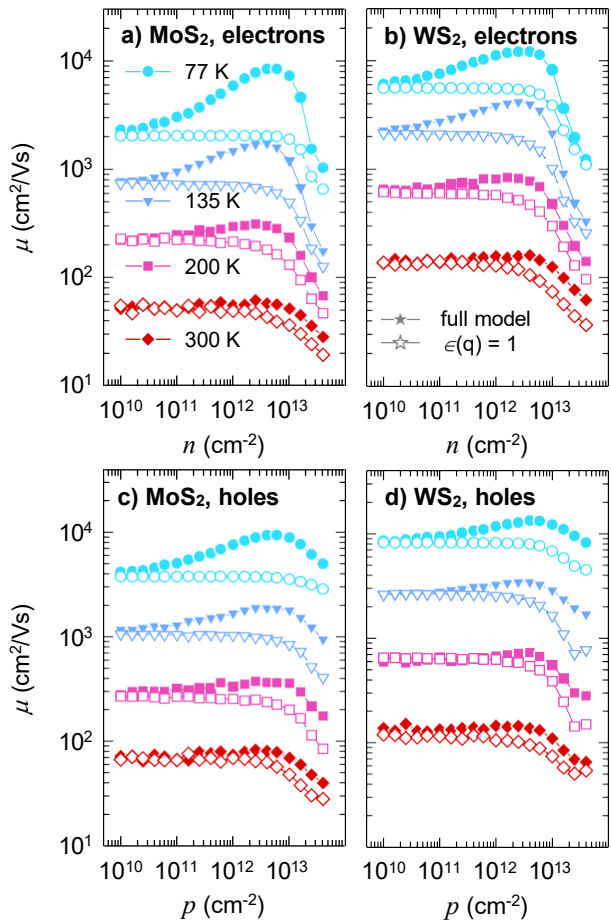

**Figure 1.** Electron (a and b) and hole (d and d) mobility dependence as a function of carrier density at a low electric field for (a, c) MoS$_2$ and (c, d) WS$_2$. Solid symbols stand for the results with the full model (a degenerate model that accounts for screening) while open symbols stand for the simulations without screening.

The static polarizability function decreases with temperature and thus the screening weakens as the temperature increases (see $1/\epsilon(q)^2$ in Figure S2 in the supplementary material). When $T$ increases, the Fermi-Dirac distribution function also widens, and its tail spans to more energetic states. Therefore, a greater amount of carriers show larger energies and experience scattering with long $q$ transition vectors, which additionally weakens the global effect of screening.

Figure 2 presents electron occupations of K and Q valleys as a function of the electron concentration at different temperatures. The hole occupations of K and Γ valleys are not shown in the graphs, being practically 100% for the K valleys in all the hole concentration range under study and regardless of the lattice temperature. This is a consequence of a larger difference in the valleys potential energies within the valence band. For the electrons on the other hand, when lattice temperature increases, the kinetic energy of electrons rises too, leading to an increased probability of electrons to transfer into the upper Q valleys which have a heavier effective electron mass (see Table 1). At low electron concentrations, lattice temperature is the leading parameter to determine the occupation of the upper valleys. Practically all electrons (near 100%) remain in the K valley until $n$ reaches sufficiently large values in the range of $10^{12} - 10^{13}$ cm$^{-2}$. At larger electron concentrations, the occupation of the Q valleys increases significantly, indicating that the Fermi level is approaching the potential energy of those upper valleys. Besides, as lattice temperature increases, the increase in the occupancy of Q valleys is attained even at low electron concentrations as a result of the broadened distribution tails spanning to the bottom of these valleys minima. A larger change in the occupancy of the primary valleys is also observed when electron concentration increases from low to large in WS$_2$, because of its reduced density of states related to the smaller effective mass in its K valleys. Note that the electron occupation of upper valleys in MoS$_2$ and WS$_2$ will induce further scattering modes involving transitions between K and Q valleys, and also within Q valleys.

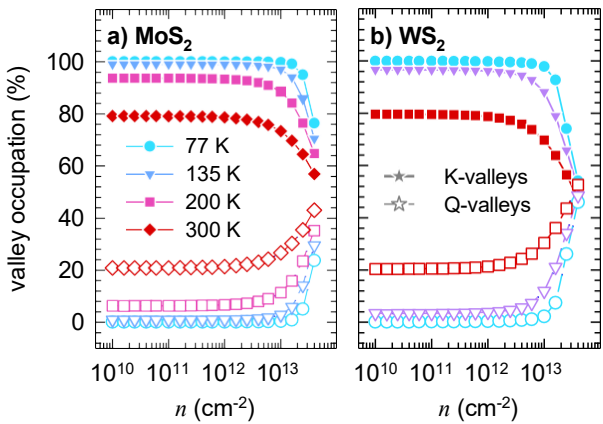

**Figure 2.** Percentage of electrons in K (full symbols) and Q valleys as a function of the electronic concentration at a low electric field (0.5 kV/cm) for (a) MoS$_2$ and (b) WS$_2$ at four different temperatures.

Figure 3 shows an inverse momentum relaxation time as a function of carrier concentration for MoS$_2$ and WS$_2$ at four different temperatures (77 K and 300 K) for electrons and holes. The inverse momentum relaxation time is computed from the monitoring of a total number of scatterings suffered by carriers at a low electric field (0.5 kVcm). The result without considering screening is also shown for comparison. Besides, in the case of electrons, the contributions of scattering mechanisms including phonon scattering between K-

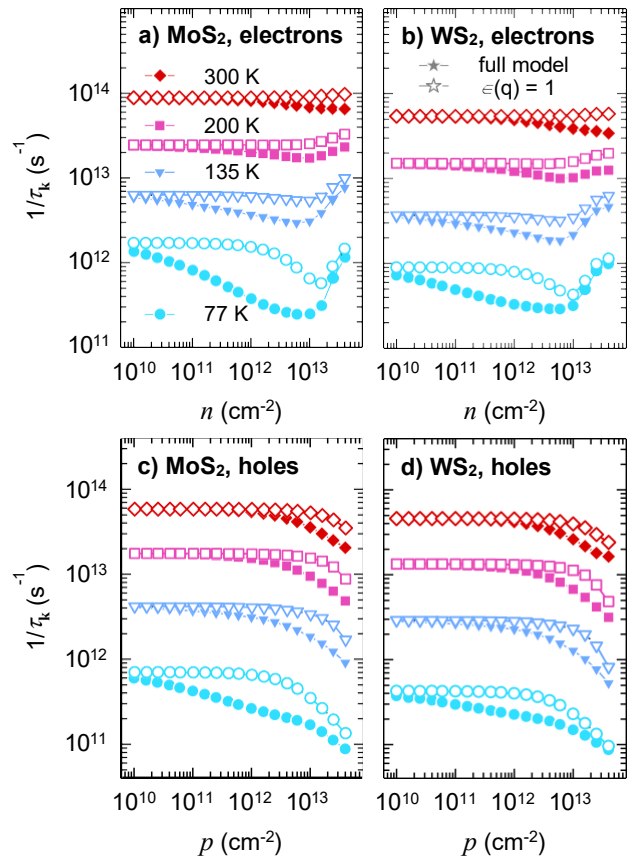

**Figure 3.** Momentum relaxation rates as a function of the electron (a and b) and hole (c and d) concentrations at low electric field (0.5 kV/cm) for (a and c) MoS$_2$ and (b and d) WS$_2$ at four different temperatures. Solid symbols stand for the results with the full model including the screening while open symbols stand for the simulations excluding the screening.

K, K-Q, and Q-Q valleys, and the scattering with surface polar phonons in the K and Q valleys can be examined in Figure 4. At a low lattice temperature, a progressive reduction in $1/\tau_k$ is observed in both materials for electrons and holes, reaching the largest difference at intermediate concentrations around $10^{13}$ cm$^{-2}$ when the transport model excludes the screening, after which $1/\tau_k$ tends to increase for the electrons and thus the differences diminish. The dominant scattering mechanism for this behaviour is the K-K intrinsic phonon-assisted transition. The transition dominates the scattering at low $T$ (see Figure 4) and is strongly affected by screening, thus explaining the mobility enhancement. On the other hand, at large carrier concentrations, the aggregated K-Q and Q-Q scattering modes become more relevant as the occupation of the Q valleys grows, thus increasing $1/\tau_k$ at the largest carrier concentrations, as reported in Figure 2. This is the main reason that explains relatively small gain in mobility at room temperature in comparison to the results reported in [18]. In the case of holes, this increase at high concentration values does not occur, being related to the low population of the $\Gamma$ valleys in the whole concentration range.

At room temperature, independently of the type of carrier, the effect of screening on scattering is less important. The electron-SPP interactions within the K valleys are the dominant scattering up to about $10^{13}$ cm$^{-2}$, while intrinsic K-K scatterings are less relevant (see Figure 4). The SPP scattering involves larger phonon wavevectors with greater phonon energies in the emission/absorption process, corresponding to smaller dielectric function values, and thus reducing the screening effect. Consequently, at room temperature, screening does not provide a significant electron mobility enhancement. In addition, the scattering mechanisms in the Q valleys become more relevant than at low temperatures. In the case of holes, intrinsic K-K scatterings are the most relevant scattering mechanisms at low temperature, being hole-SPP interactions within the K valleys the dominant ones at high temperature (see Figure S4 in the supplementary material), thus explaining the reduction of the screening effect observed at 300 K in the whole concentration range.

For the analysis of the electron high field drift velocity, a electric field value of 30 kV/cm has been considered. The results for MoS$_2$

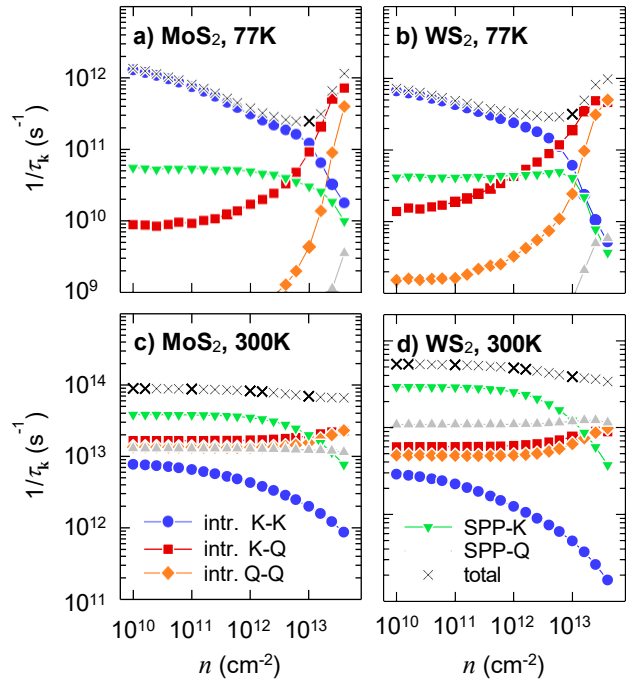

**Figure 4.** Intrinsic momentum relaxation rates as a function of the electron carrier concentration at a low electric field (0.5 kV/cm) for the electron scattering between the K-K valleys (including intra- and inter-valley transitions), the K-Q valleys, the Q-Q valleys (including intra- and inter-valley transitions), the SPP-K (the SPP interactions in the K valleys), and the SPP-Q (the SPP interactions in the Q valleys) at temperatures of (a) and (b) 77 K, and (c) and (d) 300 K.

and WS$_2$ are depicted in Figure 5 (a) and (b), respectively, as a function of the carrier concentration, at four different temperatures. The values and trends for holes are similar (see supplementary material, Figure S5). The drift electron velocity is steadily decreasing as the temperature increases. Similarly to the low-field conditions, the temperature strongly influences the population distribution of the different valleys at high electric fields.

It should be noted that, despite the fact that a strong electric field makes carriers attain higher kinetic energies, the relative percentage distribution of carriers between primary and secondary valleys is similar to that observed at low fields. The increase in kinetic energy of electrons provokes the activation of SPP emissions, so the SPP interactions in the K-valleys are now a dominant scattering mechanism along the whole temperature range under consideration (see the supplementary material), acting as efficient pathways for energy relaxation. Carriers reach the upper valleys less frequently in samples with an underlying substrate when compared to suspended (substrate-free) TMD layers [33].

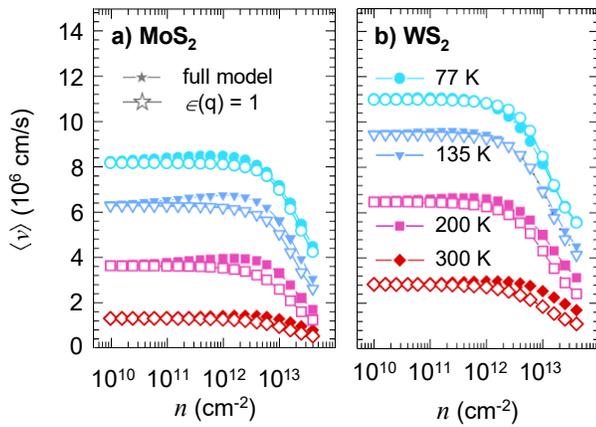

**Figure 5.** Electron drift velocity as a function of the carrier concentration at a high electric field of 30 kV/cm assuming four indicated lattice temperatures for (a) MoS$_2$ and (b) WS$_2$.

The analysis of the dependence of the drift velocity against carrier concentration indicates that the drift electron velocity is not affected by the screening under high electric fields so strongly as the electron mobility is at low electric field. Electrons under the influence of a high electric field gain higher kinetic energy, thus making the SPP scattering the dominant one, as already noted. In addition, the overall large phonon wavevectors involved in electron interactions at high electric field prevent a strong screening action. In MoS$_2$, the screening has influence at intermediate temperatures and intermediate concentrations, but in WS$_2$, the effect is observed mainly at room temperature and large carrier concentrations. This can be explained by the differences between both materials in electron effective masses of conduction valleys and by a relatively more important intrinsic phonon transitions in MoS$_2$ as compared to WS$_2$.

## 4. Conclusions

The effect of free carrier screening and degeneracy on the electronic transport properties of 2D TMD materials on a SiO$_2$ substrate has been analysed by an in-house ensemble Monte Carlo simulator. We focused on two of the most relevant TMDs: molybdenum disulphide (MoS$_2$) and tungsten disulphide (WS$_2$).

A strong non-monotonic dependence of the extracted low-field mobility with the carrier concentration has been observed at the lowest temperature under study. Indeed, the highest mobility has been reached at the lowest sampled temperature ($T$ = 77 K) with $n \approx 6\times10^{12}$ cm$^{-2}$ for MoS$_2$, and $n \approx 4\times10^{12}$ cm$^{-2}$ for WS$_2$, with values of $\sim$ 8400 cm$^2$/Vs and $\sim$ 12040 cm$^2$/Vs, respectively, that represent over a 4-fold and 2-fold increases in mobility. As for holes, maximum mobilities are attained at the same sampled temperature, reaching 9320 cm$^2$/Vs and 13290 cm$^2$/Vs for MoS$_2$ and WS$_2$ respectively, being the enhancement relative to the non-degenerate case less remarkable than for electrons. At intermediate carrier concentrations, the progressive increase of electron mobility up to maximum values stems from the effect of screening on intrinsic scattering mechanisms in the K valleys. Therefore, a complete consideration of screening (including intrinsic phonons) in carrier transport model is mandatory. At larger electron concentrations, the ob-served drop in their mobility comes as the result of the increasing proportion of electrons reaching the upper Q valleys (with a heavier effective mass) due to degeneracy. The increasing occupation of the Q valleys also leads to the onset of additional electron scattering mechanisms (SPP-K, SPP-Q) that contribute to transport degradation. In the case of holes, the impact of secondary valleys ($\Gamma$) in carrier transport was found to be marginal within the simulation conditions, due to the minimal upper valley occupation stemming from a larger energy separation.

The electron drift velocity at a high electric field is strongly influenced by the SPP scattering in the K valleys, which becomes dominant in that regime, acting also as a very effective energy relaxation mechanism. This translates into a much weaker dependence of the upper Q valley occupation on the electric field in comparison with suspended (free standing) TMDs [33]. Moreover, we have demonstrated that the screening effect at these high electric fields is less important than at low fields due to large phonon wavevectors involved in the SPP interactions, that imply a smaller effective dielectric function.